\documentclass[a4paper, 12pt]{article}

\usepackage{changepage}

\usepackage[utf8]{inputenc}
\usepackage[T1]{fontenc}

\usepackage[]{graphicx}
\usepackage{caption}
\usepackage{subcaption}
\usepackage{mathtools}

\usepackage{epsf,amsmath,bbold,amsfonts,stmaryrd}
\usepackage{mathrsfs}
\usepackage{tikz}
\usetikzlibrary{arrows,matrix,positioning}
\usepackage{appendix}
\usepackage{amssymb}
\usepackage{float}
\usepackage{color} 
\usepackage{dsfont}
\usepackage{bm}
\usepackage[colorlinks]{hyperref}
\hypersetup{pageanchor=false,citecolor=red,urlcolor=red}
\usepackage{indentfirst}
\usepackage{slashed}
\usepackage[numbers,square,comma,compress,merge]{natbib}

\usepackage[skip=2pt,font=small]{caption}

\allowdisplaybreaks

\def\a{\alpha}
\def\b{\beta}
\def\c{\chi}

\def\d{\delta}

\def\f{\frac}
\def\g{\gamma}

\def\l{\left}

\def\m{\mu}
\def\n{\nu}

\def\p{\partial}

\def\r{\right}

\def\x{\xi}

\def\z{\zeta}

\def\be{\begin{equation}}
\def\ee{\end{equation}}

\def\bes{\begin{equation*}}
\def\ees{\end{equation*}}

\def\bea{\begin{eqnarray}}
\def\eea{\end{eqnarray}}

\def\ba{\begin{array}}
\def\ea{\end{array}}

\def\bc{\begin{center}}
\def\ec{\end{center}}

\def\bl{\begin{flushleft}}
\def\el{\end{flushleft}}

\def\br{\begin{flushright}}
\def\er{\end{flushright}}

\def\bi{\begin{itemize}}
\def\ei{\end{itemize}}

\def\bt{\begin{tabular}}
\def\et{\end{tabular}}

\newtheorem{question}{Question}
\def\bq{\begin{question}}
\def\eq{\end{question}}

\newtheorem{definition}{Def}
\def\bd{\begin{definition}}
\def\ed{\end{definition}}

\newtheorem{answer}{Answer}
\def\ban{\begin{answer}}
\def\ean{\end{answer}}

\newtheorem{possibleanswer}{Possible answer}
\def\bpa{\begin{possibleanswer}\normalfont}
\def\epa{\end{possibleanswer}}

\newtheorem{theorem}{Theorem}
\def\bth{\begin{theorem}}
\def\eth{\end{theorem}}

\usepackage{esvect}

\usepackage{slashed}

\usepackage{xspace}
\newcommand*{\ie}{i.e., }
\newcommand*{\eg}{e.g., }

\newcommand*{\Eq}{Eq.\@\xspace}

\usepackage{xifthen}
\newcommand*\diff{\mathrm{d}} 
\newcommand*\ldiff[2][]{ \ifthenelse{\isempty{#1}}{ \diff #2}{\diff^#1#2} \,} 
\let\limitint\int 
\renewcommand{\int}{\limitint \!} 

\newsavebox\myboxA
\newsavebox\myboxB
\newlength\mylenA

\newcommand*\xoverline[2][0.75]{%
    \sbox{\myboxA}{$\m@th#2$}%
    \setbox\myboxB\null
    \ht\myboxB=\ht\myboxA%
    \dp\myboxB=\dp\myboxA%
    \wd\myboxB=#1\wd\myboxA
    \sbox\myboxB{$\m@th\overline{\copy\myboxB}$}
    \setlength\mylenA{\the\wd\myboxA}
    \addtolength\mylenA{-\the\wd\myboxB}%
    \ifdim\wd\myboxB<\wd\myboxA%
       \rlap{\hskip 0.5\mylenA\usebox\myboxB}{\usebox\myboxA}%
    \else
        \hskip -0.5\mylenA\rlap{\usebox\myboxA}{\hskip 0.5\mylenA\usebox\myboxB}%
    \fi}
\makeatother

\begin{document}

\begin{titlepage}

\vspace*{-2.5cm}
\begin{adjustwidth}{}{-.45cm}
\br
{
\begin{tabular}{@{}l@{}}
\small LMU--ASC~28-21 \\
 \small FTPI--MINN--21--14 \\
 \small UMN--TH--4022/21
 \end{tabular}
 }
\er
\end{adjustwidth}

\vspace*{.5cm}

\begin{center}
    \bf \Large{Scale and Weyl Invariance in Einstein-Cartan Gravity}
\end{center}

\begin{adjustwidth}{-1.3cm}{-.7cm}
\begin{center}
\textsc{Georgios K. Karananas,$^{\star}$~Mikhail Shaposhnikov,$^\dagger$\\
Andrey Shkerin,$^{\ddagger}$~Sebastian Zell\,$^\dagger$}
\end{center}
\end{adjustwidth}

\begin{center}
\it {$^\star$Arnold Sommerfeld Center\\
Ludwig-Maximilians-Universit\"at M\"unchen\\
Theresienstra{\ss}e 37, 80333 M\"unchen, Germany\\
\vspace{.4cm}
$^\dagger$Institute of Physics\\
Laboratory of Particle Physics and Cosmology\\
\'Ecole Polytechnique F\'ed\'erale de Lausanne (EPFL)\\ 
CH-1015, Lausanne, Switzerland\\
\vspace{.4cm}
$\ddagger$William I. Fine Theoretical Physics Institute\\
School of Physics and Astronomy\\
University of Minnesota \\
Minneapolis, MN 55455, USA
}
\end{center}

\begin{center}
\small
\texttt{georgios.karananas@physik.uni-muenchen.de}\\
\texttt{mikhail.shaposhnikov@epfl.ch}\\
\texttt{ashkerin@umn.edu}\\
\texttt{sebastian.zell@epfl.ch}
\end{center}

\vspace{1cm}

\begin{abstract}
We show how Einstein-Cartan gravity can accommodate both global scale and local scale (Weyl) invariance. To this end, we construct a wide class of models with nonpropagaing torsion and a nonminimally coupled scalar field. In phenomenological applications the scalar field is associated with the Higgs boson. For global scale invariance, an additional field --- dilaton --- is needed to make the theory phenomenologically viable. In the case of the Weyl symmetry, the dilaton is spurious and the theory reduces to a sub-class of one-field models.  In both scenarios of scale invariance, we derive an equivalent metric theory and discuss possible implications for phenomenology.
\end{abstract}

\end{titlepage}

\section{Introduction and motivation}

Einstein's theory of General Relativity (GR) exists in different versions. Along with the most commonly used metric approach, these include the Palatini, affine, and teleparallel formulations, as well as Einstein-Cartan gravity. All these theories are equivalent in the absence of matter. A priori, this puts all of them on the same footing. Generically, however, this degeneracy disappears once mater is included. Then the various versions of gravity no longer lead to identical predictions, and, consequently, phenomenology depends on the choice of formulation. Recently, we have undertaken an effort to quantify these differences between the distinct incarnations of GR~\cite{matter_matters}. There, our focus was on the Einstein-Cartan (EC) formulation, which encompasses the metric and Palatini gravity as special cases. We first proposed a set of criteria for systematically constructing a theory of matter coupled to EC gravity and then we included in the action all terms that fulfill these criteria. Our results generalize numerous previous studies~\cite{gr-qc/0505081, hep-th/0507253,Alexandrov:2008iy, 0807.2652, 0811.1998, 0902.0957, 0902.2764, Diakonov:2011fs, 1212.0585, Langvik:2020nrs,Shaposhnikov:2020frq}.

In the metric formulation of GR, one assumes a priori that torsion is absent. By allowing for non-vanishing torsion, one arrives at EC gravity. As said, the two theories are equivalent in the absence of matter. In particular, this implies that the gravitational spectrum is identical in both cases, and only consists of two degrees of freedom of the massless graviton. Hence, torsion does not propagate in the EC gravity, and the theory can be solved for it. By plugging the solution back into the action, one derives an equivalent theory in the metric formulation in which torsion is replaced by a specific set of higher-dimensional operators. The suppression scale of these operators can, in principle, be (much) lower than the Planck scale. In this case, the effects of torsion may become visible at energies relevant for cosmology or even in particle physics experiments. This in turn can put constraints on the parameters of the theory. Possible implications of EC gravity for phenomenology have been studied, e.g., in~\cite{Shaposhnikov:2020aen} with regard to the production of fermionic dark matter and in~\cite{Langvik:2020nrs, Shaposhnikov:2020gts} in the context of Higgs inflation. 

The goal of the present paper is to liberate from explicit mass scales the  graviscalar part of the EC model constructed in~\cite{matter_matters}. We will do so in two different ways. On the one hand, we will make the theory invariant under global dilatations, and on the other hand under local scale transformations. We will refer to the latter case as Weyl symmetry. In the following, we shall discuss why generalizing the theory in these distinct, nevertheless interconnected, ways may be potentially interesting. Note that here we will restrict ourselves to the classical aspects of the theory, leaving the discussion of quantum effects to future work.

Let us begin with the case of global dilatations. That this symmetry (as well as the closely related conformal invariance in flat spacetime) can have rich ramifications for both cosmology and particle physics was realized many years ago~\cite{Englert:1976ep,Wetterich:1983bi,Wetterich:1987fk,Wetterich:1987fm,Dehnen:1992jc,Wetterich:1994bg,Cervantes-Cota:1995ehs,Bardeen:1995kv} and since then there has been an increasingly rising interest in this direction. For a non-exhaustive list of references see~\cite{Meissner:2006zh,Foot:2007iy,Shaposhnikov:2008xb,
Shaposhnikov:2008xi,Shaposhnikov:2008ar,Shaposhnikov:2009nk,
GarciaBellido:2011de,Blas:2011ac,GarciaBellido:2012zu,Bezrukov:2012hx, Bars:2013yba,
Monin:2013gea,Armillis:2013wya,Tavares:2013dga,Gretsch:2013ooa,
Khoze:2013uia,Csaki:2014bua,Rubio:2014wta,Ghilencea:2015mza,Karam:2015jta,
Karananas:2015eha,Karananas:2015ioa,
DiVecchia:2015jaq,Lucat:2016eze,Trashorras:2016azl, Karananas:2016grc,
Ferreira:2016vsc,Bianchi:2016viy,
Karananas:2016kyt,Karam:2016rsz,Ferreira:2016wem,Ferreira:2016kxi,Ghilencea:2016dsl,
Shkerin:2016ssc,Karananas:2017mxm,Rubio:2017gty,
DiVecchia:2017uqn,Guerrieri:2017ujb,Tokareva:2017nng,Gillioz:2017ooc,
Karananas:2017zrg,Lucat:2017wtu,Casas:2017wjh,
Ferreira:2018itt,Ferreira:2018qss,Shaposhnikov:2018xkv,Lucat:2018slu,
Shaposhnikov:2018jag,Burrage:2018dvt,Loebbert:2018xsd,SravanKumar:2018tgk,Lalak:2018bow,
Gorbunov:2018llf,Iosifidis:2018zwo,Quiros:2018ryt,Casas:2018fum,Shkerin:2019mmu,Herrero-Valea:2019hde,Karananas:2019fox,Rubio:2020zht,Karananas:2020qkp,Hill:2020oaj}. In the setup we have in mind, the Higgs field, in addition to its significance for the Standard Model (SM), plays a central part in cosmology. Namely, it is responsible for the inflationary stage of the Universe. In the  metrical setup, this can be achieved in the context of the Higgs-dilaton model~\cite{Shaposhnikov:2008xb}. To achieve global scale invariance and to yield an acceptable cosmological and particle physics phenomenology, the model necessitates the presence of yet another scalar field, the dilaton, a singlet under the SM symmetries. This has two roles. On the one hand, being the order parameter for dilatations, it induces all scales at the classical level. On the other hand, it can act as dynamical dark energy accounting for the present-day accelerated  expansion of the Universe.\footnote{Of course, this can instead be attributed to a cosmological constant. In the scale-invariant approach it is associated with the quartic self-interaction of the dilaton~\cite{GarciaBellido:2011de}.}~This comes into fruition by restricting the spacetime symmetry to the volume-preserving subgroup of diffeomorphisms \cite{vanderBij:1981ym, Kreuzer:1989ec}. In the prototype Higgs-dilaton model, unimodular gravity \cite{Unruh:1988in, Henneaux:1989zc} was employed---a special case of the restricted general coordinate transformations. Interestingly, being the scale-donor, the dilaton leaves its imprints in the inflationary observables, implying nontrivial links  between the very early and late  Universe. Vaguely speaking, the deviation of the inflationary spectrum from exact scale invariance is closely tied with the deviation of the dark energy from a (cosmological) constant~\cite{GarciaBellido:2011de}. 

In addition to its vast applications to cosmological physics, scale invariance could also be of relevance for the SM finetuning issues~\cite{Wetterich:1983bi,Bardeen:1995kv}. This requires that the symmetry, in addition to being spontaneously broken, must also be exact at the quantum level. When the no-scale approach is combined with the assumption of no new particle states between the electroweak and Planck energies~\cite{Shaposhnikov:2007nj}, the Higgs mass is insensitive to loop effects~\cite{Shaposhnikov:2008xi}. Although this takes care of the stability part of the hierarchy puzzle, it does not explain the 34 orders of magnitude difference between the value of the electroweak and Planck scales. This may well be attributed to the nonperturbative scale-splitting mechanism introduced in~\cite{Shaposhnikov:2018xkv} and further studied in~\cite{Shaposhnikov:2018jag,Shkerin:2019mmu,Shaposhnikov:2020geh,Karananas:2020qkp}; scale-free models constitute a natural playground for generating hierarchies this way. 
Notice also that when symmetry breaking vacua (flat directions) exist and the system evolves along them, the vacuum energy is zero; see, e.g.,~\cite{Amit:1984ri,Einhorn:1985wp,Rabinovici:1987tf,Shaposhnikov:2008xi,Karananas:2019fox}. This is irrespective of the fact that mass scales are present. In turn, spontaneous scale (or conformal) symmetry breaking may have implications for the cosmological constant problem as well.    

We now move to the Weyl symmetry. Although Weyl's original theory has been around since 1918~\cite{Weyl:1918ib}, the various implications of Weyl invariance for cosmological model building have only been investigated fairly recently; see, e.g.,~\cite{Kallosh:2013hoa,Kallosh:2013daa,Oda:2018zth,Barnaveli:2018dxo,Edery:2019txq,Ferreira:2019zzx,Lin:2020phk,Hobson:2020doi,Tang:2020ovf}. Interestingly, there is more than one way to gauge scale symmetry. It is obvious that one can localize it at the expense of a new dynamical compensating gauge field. This, however, would add extra degrees of freedom in the theory. Here we wish to refrain from doing so, but rather follow a more minimalistic approach on which we elaborate now. Weyl transformations affect nontrivially the Riemannian curvatures and torsion, and the latter transform in an inhomogeneous manner. This implies that, at least in principle, it is possible to make the theory invariant under gauged scale transformations without introducing new degrees of freedom,\footnote{We are aware of only a handful of cases where this happens:~{\emph{i)}}~Einstein-Cartan gravity, where the full Poincar\'e group is gauged with the connection not being independent of the vielbein;~{\emph{ii)}}~the conformally coupled scalar field~\cite{Iorio:1996ad} and its higher-derivative generalizations~\cite{Karananas:2015eha};~{\emph{iii)}}~Galilean- and Lifshitz- invariant theories~\cite{Karananas:2016hrm};~{\emph{iv)}}~the three-form gauging of the shift symmetry of axion-like fields~\cite{Dvali:2005an}.} but rather by employing quantities of geometric origin that in any case are already present. Particularly in the EC framework, the torsion vector mimics the behavior of a Weyl gauge field, thus it can be employed to compensate for the non-invariance of the Ricci scalar and the derivatives of the fields in the action. To the best of our knowledge, the fact that the Weyl vector may be torsion in disguise was pointed out for the first time  by Nieh and Yan~\cite{Nieh:1981xk}, Dereli \& Tucker~\cite{Dereli:1982xb}, and  Yu.~Obukhov~\cite{Obukhov:1982zn}; see also the closely  related work~\cite{Nieh:1982nb}.

Let us briefly note that going beyond the EC formulation of GR, one can construct a vector out of nonmetricity---with the  appropriate transformation properties---to play the role of the Weyl field~\cite{Ghilencea:2018thl,*Ghilencea:2018dqd,*Ghilencea:2019jux,*Ghilencea:2019rqj,*Ghilencea:2020piz,*Ghilencea:2020rxc,*Ghilencea:2021lpa}. 
The logic in these approaches is different from ours, in that although nonmetricity is of geometrical origin it is taken to be dynamical from the onset. As expected, this can only be achieved by working with an action that contains curvature-squared terms. This actually serves as a perfect example for what we discussed in~\cite{matter_matters}. The admissible terms have to be chosen carefully to avoid pathologies: for instance, a term proportional to the (Weyl tensor)$^2$ is not included, since it introduces a ghostly massive spin-2 field in the spectrum. 

It should be made clear that in our approach, torsion is always nonpropagating, so Weyl invariance is actually nothing more than a ``book-keeping'' device: it allows us to restrict the number of admissible terms in the action. This point was also stressed in~\cite{Jackiw:2014koa}. Let us also mention that in the Weyl-invariant embedding, the gauge freedom eliminates completely  one scalar degree of freedom. This means that since our starting point is the EC counterpart of the Higgs-dilaton model, there will be 2+1 dynamical degrees of freedom in the graviscalar spectrum. 

The paper is organized as follows. In Sec.~\ref{sec:recap}, we recap the results of~\cite{matter_matters} concerning the graviscalar piece of the nonminimally coupled EC theory. 
In Sec.~\ref{sec:scal_inv}, we construct the globally scale invariant generalization of the theory. We explicitly demonstrate why it is necessary to introduce the dilaton. We obtain the EC counterpart of the metrical Higgs-dilaton theory. We diagonalize the Einstein-frame kinetic sector of the theory and identify the physical dilaton, i.e. the Nambu-Goldstone field of the nonlinearly realized dilatations. In Sec.~\ref{sec:weyl_inv}, we construct the Weyl invariant version of the theory and discuss in details the resulting constraints on the action. 
 For completeness we also briefly comment on the inclusion of fermionic matter in both the scale- and Weyl- invariant settings.
Our conclusions are presented in Sec.~\ref{sec:conclusions}.  

\paragraph{Conventions.} We work exclusively in four spacetime dimensions. Greek  and capital Latin letters are used for spacetime and Lorentz indices, respectively. The signature of both the spacetime $g_{\a\b}$ and Minkowski $\eta_{AB}$ metrics is mostly plus. As usual, the tetrad/vierbein is denoted with $e_\m^A$ and the spin connection by $\omega_\m^{AB}$.  We set $c=\hbar=1$.

\section{EC gravity with nonminimally coupled scalar field}
\label{sec:recap}

\subsection{Criteria for constructing the action}
\label{ssec:criteria}
In this section, we review the criteria developed in \cite{matter_matters} for constructing an action of the SM coupled to gravity in the EC formulation. Our basic requirement is that in the flat limit the SM is recovered, while in the absence of matter one obtains a theory of gravity that is equivalent to GR in the metric formulation. In particular, there should be no additional propagating degrees of freedom as compared to the ones of the SM plus the massless graviton. In order to ensure that these conditions are fulfilled, we put forward the following criteria in~\cite{matter_matters}
\begin{itemize}
   \item[\emph{i)}] The purely gravitational part of the action should solely contain operators of mass dimension not greater than 2.
   \item[\emph{ii)}] In the flat spacetime limit, \ie for $e_\m ^A  =  \d_\m^A,~\omega_\m^{AB} = 0$, the matter Lagrangian should be renormalizable.
   \item[\emph{iii)}] The coupling of matter to gravity should only happen through operators of mass dimension not greater than 4. 
\end{itemize}
As explained in~\cite{matter_matters}, condition~\emph{i)} is sufficient, although not necessary, to make sure that the two polarizations of a massless spin-2 field are the only propagating degrees of freedom in gravity. Furthermore, criterion~\emph{ii)} is crucial for the predictiveness of our setup. After solving for torsion and plugging the result back in the action, one can obtain an equivalent theory in the metric formulation that features a specific set of higher-dimensional operators. Without criterion~\emph{ii)}, one could have added from the beginning all kinds of higher-dimensional operators, and our procedure would be meaningless. Finally, since both the pure matter and pure gravity sectors only contain operators of mass dimension not greater than 4, it is natural to impose that all terms in the action exhibit this property, which leads to condition~\emph{iii)}. However, it is possible to relax this last criterion. The implications of imposing (or not imposing) it will be made explicit shortly.

\subsection{Equivalent metric theory}

From now on we solely focus on EC gravity coupled nonminimally to a real scalar field, which we denote by $h$. The generalization to a complex field, as well as the inclusion of gauge fields and fermions, are discussed in~\cite{matter_matters}. At this point, we shall only impose the conditions~\emph{i)} and~\emph{ii)} discussed above, and this leads to the action\,\footnote{Unless otherwise stated, summation over repeated indices is assumed.}
\begin{equation}
\begin{aligned}
\label{eq:action_real_scalar_F}
S_{\rm{gr} +h} &= \int \diff^4 x \sqrt{g} \Bigg [ M_P^2\Omega^2 F +M_P^2\tilde{\Omega}^2 \tilde F - \frac {(\p_\m h)^2}{2} -U + v^\mu \partial_\mu z_v + a^\mu \partial_\mu z_a  \\
&\qquad+ \f{M_P^2}{2}\Big(\mathbf{g}_{vv} v_\mu v^\mu  + 2\mathbf{g}_{va}v_\mu a^\mu + \mathbf{g}_{aa}	a_\mu a^\mu +\mathbf{g}_{\tau\tau}\tau_{\a\b\g} \tau^{\a\b\g} + \tilde{\mathbf{g}}_{\tau\tau} \epsilon^{\m \n \rho \sigma} \tau_{\lambda\m\n} \tau^\lambda_{~\rho\sigma}\Big) \Bigg]\ ,
\end{aligned}
\end{equation}
where we defined $g = - \det(g_{\mu\nu})$. Let us explain the various terms appearing in this expression. We denoted with $F$ and $\tilde F$ the parity-even Einstein-Hilbert  and parity-odd Holst scalars \cite{Hojman:1980kv, Nelson:1980ph, Castellani:1991et, Holst:1995pc}. In terms of the curvature 
\be
\label{eq:curv_def}
F_{\m\n}^{AB} = \p_\m \omega_\n^{AB} -\p_\n \omega_\m^{AB}+\omega^A_{\m C}\omega^{CB}_\n - \omega^A_{\n C}\omega^{CB}_\m \ , 
\ee
these respectively read
\be
\label{eq:f_def}
F \equiv \frac {1}{8\sqrt g} \epsilon_{ABCD}\epsilon^{\m\n\rho\sigma}F^{AB}_{\m\n}e^C_\rho e^D_\sigma \ ,~~~\text{and}~~~\tilde F \equiv \f {1}{\sqrt g}\epsilon^{\m\n\rho\sigma}e_{\rho C}e_{\sigma D}F_{\m\n}^{CD}  \ .
\ee
On the other hand,
\be 
\label{eq:tors_all}
v_\m = T^\n_{~\m\n} \ ,~~~a_\m = \epsilon_{\m\n\rho\sigma}T^{\n\rho\sigma} \ ,~~~\tau_{\m\n\rho} =\frac 2 3 \l( T_{\m\n\rho} -v_{[\n} g_{\rho]\m} - T_{[\n\rho]\m} \r) \ ,
\ee
are the irreducible components of the torsion tensor, with the latter given by 
\be
\label{eq:tors_def}
T_{\m\n\rho}=e_{\mu A} T^A_{\n\rho}\ ,~~~T_{\m\n}^A = \p_\m e_\n^A -\p_\n e_\m^A+ \omega^A _{\m B}e^B_\n -\omega^A _{\n B}e^B_\m  \ .
\ee
Finally, $\Omega^2$, $\tilde{\Omega}^2$, $U$, $z_i$ (which can stand for $z_v$ and $z_a$) as well as $\mathbf{g}_{ij}$ (which can stand for $\mathbf{g}_{vv}$, $\mathbf{g}_{va}$, $\mathbf{g}_{aa}$, $\mathbf{g}_{\tau\tau}$ and $\tilde{\mathbf{g}}_{\tau\tau}$) are various coefficient functions of the field $h$. The first two of them represent the nonminimal couplings of $h$ to the Ricci curvature and to the Holst term, the third is the scalar field's potential, and the rest are couplings to the torsion components. 

A priori, the functions $\Omega^2$, $\tilde{\Omega}^2$, $U$, $z_i$ and $\mathbf{g}_{ij}$ can be arbitrary. If criterion~\emph{iii)} is imposed, however, this restricts them to be at most quadratic in the field $h$. Additionally requiring invariance under $h\mapsto -h$, we can then write
\begin{equation}
\label{eq:coeff_functions}
    \Omega^2=1+\frac{\xi_h h^2}{M_P^2} \;, ~~~ \tilde{\Omega}^2 = \frac{1}{4\bar{\gamma}}\left(1+\frac{\zeta_h h^2}{M_P^2}\right) \;, ~~~ z_i = \zeta_i h^2\;, ~~~ \mathbf{g}_{ij}= c_{ij} \l(1+\frac{\x_{ij}h^2}{M_P^2}\r) \;,
\end{equation}
where no summation over repeated indices is implied. Now $\xi_h$, $\bar{\gamma}$, $\zeta_h$, $\zeta_i$, $c_{ij}$ and $\xi_{ij}$ are real numbers; in the first of them we recognize the standard nonminimal coupling of $h$ to curvature, and in the second one the so-called Barbero-Immirzi parameter \cite{Immirzi:1996dr,Immirzi:1996di}. 

It is important to note that the connection and, hence, the Einstein-Hilbert and Holst terms can be split into torsion-free and torsionful pieces. Moreover, the former are zero for the Holst term. The torsionful part has a form which is already accounted for in Eq.~(\ref{eq:action_real_scalar_F}). Hence, the graviscalar action can be expressed in the following form
\be
\begin{aligned}
\label{eq:action_real_scalar}
S_{\rm{gr}+h} &= \int \diff^4 x \sqrt{g}\Bigg[\frac{M_P^2}{2}\Omega^2\mathring{R} - \frac {(\p_\m h)^2}{2} - U + v^\mu \partial_\mu Z_v + a^\mu \partial_\mu Z_a \\
&+ \f{M_P^2}{2}\Big(G_{vv} v_\mu v^\mu  + 2G_{va}v_\mu a^\mu + G_{aa}	a_\mu a^\mu 
 +G_{\tau\tau}\tau_{\a\b\g} \tau^{\a\b\g}+ \tilde{G}_{\tau\tau} \epsilon^{\m \n \rho \sigma} \tau_{\lambda\m\n} \tau^\lambda_{~\rho\sigma}\Big) \Bigg]\ ,
\end{aligned}
\ee
with $\mathring R$ the Ricci scalar, and the coefficient functions in Eqs.~(\ref{eq:action_real_scalar_F}) and (\ref{eq:action_real_scalar}) are related via
\bea
&\quad\quad z_v =Z_v+M_P^2\Omega^2 \ ,~~~z_a =Z_a-M_P^2\tilde{\Omega}^2\ ,~~~\mathbf{g}_{vv}=G_{vv}+\dfrac{2\Omega^2}{3} \ ,
\\
&\mathbf{g}_{va}=G_{va}-\dfrac{2\tilde{\Omega}^2}{3}  \ ,~~\mathbf{g}_{aa}=G_{aa}-\dfrac{ \Omega^2}{24}  \ ,~~\mathbf{g}_{\tau\tau}=G_{\tau\tau}-\dfrac{\Omega^2}{2}  \ ,~~\tilde{\mathbf{g}}_{\tau\tau}=\tilde G_{\tau\tau}+\tilde{\Omega}^2  \ . \label{eq:g_G}
\eea
We notice that there is a degeneracy in the coefficient functions in~\Eq (\ref{eq:action_real_scalar_F}); in particular, $\tilde{\Omega}^2$ can be absorbed by redefining the other functions. If condition~\emph{iii)} is imposed, and, consequently, the coefficient functions are of the form~(\ref{eq:coeff_functions}), the theory~\eqref{eq:action_real_scalar_F} (or, equivalently,~\eqref{eq:action_real_scalar}) contains a finite number of free independent parameters. In addition to the Planck scale $M_P$, these are $9$ dimensionless coupling constants. Here we did not count the parameters contained in the two functions $\mathbf{g}_{\tau\tau}$ and $\tilde{\mathbf{g}}_{\tau\tau}$. The reason is that the tensorial part of torsion $\tau_{\m\n\rho}$ vanishes on the equations of motion (see~\cite{matter_matters}), and therefore all results are independent of $\mathbf{g}_{\tau\tau}$ and $\tilde{\mathbf{g}}_{\tau\tau}$.

Finally, we would like to bring the theory~(\ref{eq:action_real_scalar_F}) (or~(\ref{eq:action_real_scalar})) to a form in which no reference to torsion is made. As explained in~\cite{matter_matters}, we can proceed as follows. First we derive the equations of motion for the irreducible components $v_\mu$, $a_\mu$ and $\tau_{\m\n\rho}$, then we solve them and finally we plug the result back into the action. It is also convenient to rewrite the model in the Einstein frame in which the nonminimal coupling of the scalar field to gravity is absent. This can be achieved by rescaling the metric as
\be
\label{eq:Weyl_resc}
g_{\m\n} \mapsto \Omega^{-2} \tilde{g}_{\m\n} \ , 
\ee 
where we omit the tilde in the following.
When these steps are effectuated (not necessarily with this order), we obtain~\cite{matter_matters}
\be
\begin{aligned}
\label{eq:action_gen_int-out}
S_{\rm{gr}+h} &=\int \diff^4 x \sqrt{g} \Bigg[ \f {M_P^2}{2}\mathring{R} -  \f{K(h)}{2}(\p_\m h)^2 - \frac{U(h)}{\Omega^4}\Bigg] \ ,
\end{aligned}
\ee
with
\be 
\label{eq:kh}
K(h) = \frac{1}{\Omega^2} + \frac{4h^2}{M_P^2\Omega^2}\frac{G_{aa}(Z'_v)^2+G_{vv}(Z'_a)^2-2G_{va}Z'_vZ'_a}{G_{vv}G_{aa}-G_{va}^2} + \frac{24 M_P^2h^2\Omega'^2}{\Omega^2} \ .
\ee
Here prime denotes a derivative with respect to $h^2$. The first term in the kinetic function (\ref{eq:kh}) is the original kinetic term in Eq.~(\ref{eq:action_real_scalar}) to which the rescaling (\ref{eq:Weyl_resc}) is applied. The second term is of purely torsional origin, and the third contribution stems from applying the transformation \eqref{eq:Weyl_resc} to the Ricci scalar. In summary, we have brought our model to a form in which torsion is replaced by a specific set of higher-dimensional operators that modify the kinetic term of the field. Therefore, we call~\Eq \eqref{eq:action_gen_int-out} the equivalent metric theory.

\section{Scale invariance}
\label{sec:scal_inv}

Our goal is to make the theory of Sec.~\ref{sec:recap} invariant under global scale transformations. The latter act on the fields as
\be
\label{eq:trans_fields}
 e^A_\m~\mapsto~q^{-1} e^A_\m \ ,~~~\omega_\m^{AB}~\mapsto~\omega_\m^{AB} \  ,~~~h~\mapsto~q h \ , 
\ee
with $q$ a constant. This leads to the following transformation laws for the Ricci scalar and the irreducible components of torsion
\be
\label{eq:trans_obj}
\mathring R~\mapsto~q^2\mathring R\ ,~~~v_\m~\mapsto~v_\m \ , ~~~a_\m~\mapsto~a_\m \ ,~~~\tau_{\m\n\rho}~\mapsto~q^{-2}\tau_{\m\n\rho} \ .
\ee
Eqs.~(\ref{eq:trans_fields}), (\ref{eq:trans_obj}) reveal that the scale-invariant generalization of the theory in the previous section follows from dimensional analysis: one simply needs to replace the Planck mass by a scalar operator of mass dimension 1.

\subsection{The need for an additional dilaton}
\label{ssec:nogo}

The most straightforward way to proceed is to use the field $h$ that is already present in the theory. This translates into considering the following action, the EC version of induced gravity
\be
\begin{aligned}
\label{eq:action_real_scalar_SI_1}
S_h &= \int \diff^4 x \sqrt{g}\Bigg[\frac{\x_h h^2}{2}\mathring{R} - \frac {(\p_\m h)^2}{2} -\f{\lambda}{4}h^4+ \z^v_h \p_\m h^2 v^\m+  \z^a_h \p_\m h^2 a^\m  \\
&\quad\qquad+ \frac{\x_h h^2}{2}\Big(c_{vv}v_\mu v^\mu  + 2 c_{va}v_\mu a^\mu + c_{aa}	a_\mu a^\mu+c_{\tau\tau}\tau_{\a\b\g} \tau^{\a\b\g} + \tilde{c}_{\tau\tau} \epsilon^{\m \n \rho \sigma} \tau_{\lambda\m\n} \tau^\lambda_{~\rho\sigma}\Big) \Bigg]\ .
\end{aligned}
\ee
Note that there is no longer any functional freedom in the coefficients, \ie $c_{vv}$, $c_{va}$, $c_{aa}$, $c_{\tau\tau}$ and $\tilde{c}_{\tau\tau}$ are real numbers.

As before, we solve for the nondynamical torsion and move to the Einstein frame via the metric redefinition~(\ref{eq:Weyl_resc}), but the conformal factor now reads
\be
\Omega = \frac{\sqrt{\x_h} h}{M_P} \ . 
\ee 
The result is given by
\be
\begin{aligned}
\label{eq:action_real_scalar_SI_1_einst}
S_h &= \int \diff^4 x \sqrt{g}\Bigg[\frac{M_P^2}{2}\mathring{R} -\frac{1}{2}\frac{M_P^2}{\kappa}\frac {(\p_\m h)^2}{h^2} -\f{\lambda M_P^4}{4\x_h^2}\Bigg]\ ,
\end{aligned}
\ee
with the coefficient $\kappa$ in the kinetic function reading
\be
\kappa=\x_h \Bigg[1+ \f{4}{\xi_h}\l(\frac{c_{aa}\z_h^{v\,2}+c_{vv}\z_h^{a\,2}-2c_{va}\z_h^v \z_h^a}{c_{vv}c_{aa}-c_{va}^2}\r)+ 6\x_h\Bigg]^{-1} \ .
\ee
The form of~(\ref{eq:action_real_scalar_SI_1_einst}) is highly suggestive: we ended up with GR with a cosmological constant term plus a minimally coupled massless field. This corresponds to the Nambu-Goldstone (NG) mode associated with the spontaneous breaking of dilatations when passing from the original to the Einstein frame. 
That this is the case becomes even more clear once we realize that the kinetic term can be canonically normalized by introducing a new field $\rho$ related to $h$ via 
\be
h = M_P {\rm e}^\f{\sqrt{\kappa}\rho}{M_P} \ ,
\ee
which is the exponential representation of the NG boson for the non-compact scale symmetry (see also the discussion in~\cite{Csaki:2014bua}).

Exactly like the induced gravity scenario in the metric formulation~\cite{Dehnen:1992jc}, the above construction is not satisfactory for phenomenology. As far as inflationary dynamics is concerned, it gives rise to an exact de~Sitter stage that does not end, i.e. there cannot be a graceful exit. The situation is even worse once we take into account that the ``scale-donor'' is the (radial mode of the) SM Higgs, for there is tension  from the particle physics point of view as well. Being a genuine NG boson, $h$ has a shift symmetry which means that it is massless and may couple only derivatively both to itself as well as to matter. It can be easily shown that when it comes to  fermions and gauge bosons $h$ decouples completely from them, the only imprint being a rescaling by powers of $\x_h$ in the respective couplings. 

A possible way out is to allow for a small explicit breaking of the scale symmetry. This is exactly what happens in Higgs~\cite{Bezrukov:2007ep}, and in Starobinsky~\cite{Starobinsky:1980te} inflationary models~\cite{Csaki:2014bua}. In both cases, it is necessary to tilt slightly the inflationary potential by allowing for the standard Einstein-Hilbert term $\propto M_P^2R$. For field values relevant for inflation this is a subdominant, nevertheless essential, contribution that modifies nontrivially the dynamics. 
Although what we just described is perfectly acceptable and well motivated, in what follows we will take another direction. Namely, we will refrain from introducing an explicit breaking of the scale symmetry, but rather have it nonlinearly realized.

\subsection{The Higgs-dilaton model in EC gravity}
\label{ssec:HD}

It turns out that the simplest and at the same time phenomenologically viable option requires that the theory is extended by another~\emph{dynamical}~scalar degree of freedom, the dilaton $\c$,\footnote{In an abuse of language we will call $\c$ the dilaton, although the physical dilaton, \ie the NG boson associated with the breaking of the scale symmetry is a function of both $h$ and $\c$.}~which is a singlet under the SM symmetries and which transforms under  dilatations in the same way as $h$. It is now this field that induces the scales in the theory; practically, we replace $M_P\mapsto\sqrt{\x_\c}\c$ in Eq.~(\ref{eq:action_real_scalar}). 

To construct the action, we first employ criteria~\emph{i)} and~\emph{ii)} from Sec.~\ref{ssec:criteria}. Additionally, since we would like to think of $h$ as the Higgs field, we do not include mixing terms like $\partial_\mu\chi\partial^\mu h$ and $\chi h$. With these restrictions, the most general biscalar action invariant under global dilatations reads
\be
\begin{aligned}
\label{eq:action_general_unitary_scale}
S_{\rm SI} &= \int \diff^4 x \sqrt{g}\Bigg[\frac{\Omega^2}{2}\mathring{R} - \frac {(\p_\m \c)^2}{2} -  \frac {(\p_\m h)^2}{2} -U(h,\c) +  J^v_\m v^\m+  J^a_\m a^\m\\
&+ \frac{\x_\c \c^2}{2}\l( G_{vv} v_\mu v^\mu  + 2G_{va}v_\mu a^\mu + G_{aa}	a_\mu a^\mu+G_{\tau\tau}\tau_{\a\b\g} \tau^{\a\b\g} + \tilde{G}_{\tau\tau} \epsilon^{\m \n \rho \sigma} \tau_{\lambda\m\n} \tau^\lambda_{~\rho\sigma}\r)   \Bigg]\ , 
\end{aligned}
\ee
with
\be
\label{eq:pot}
U(h,\c) = \f \lambda 4 \l( h^2 -\f\alpha\lambda \c^2 \r)^2 + \beta \c^4 \ .
\ee
Phenomenologically speaking, the parameter $\alpha$ in the above expression is responsible for generating the tree-level Higgs mass, while $\beta$ is responsible for the cosmological constant. 

Although the subsequent analysis can be performed without any assumptions about the form of the coefficient functions appearing in~\Eq \eqref{eq:action_general_unitary_scale}, at this point we impose criterion~\emph{iii)} from Sec.~\ref{ssec:criteria}. This will improve the clarity of the presentation and make it easier to compare with the previous studies of the Higgs-dilaton models. We find that 
\be \label{eq:conformalFactorDilaton}
\Omega^2 = \f{\x_\c \c^2+\x_h h^2}{M_P^2} \ ,
\ee
as well as 
\begin{equation}
	 J^{v/a}_\m = \zeta^{v/a}_\c \partial_\m \c^2 + \zeta^{v/a}_h \partial_\m h^2  \ .
\end{equation}
We remark that both $\xi_\chi$ and $\alpha$, $\beta$ are constrained to be much smaller than 1~\cite{GarciaBellido:2011de}. Furthermore, scale invariance dictates that the various functions appearing in~(\ref{eq:action_general_unitary_scale}) depend on ratios of the fields, i.e. 
\be
G_{ij} \equiv G_{ij}\l(\frac{h^2}{\chi^2}\r)= c_{ij}\l(1+ \f{\x_{ij}h^2}{\x_\c \c^2} \r) \ ,
\ee
where no summation over the repeated indices is implied. This is the direct generalization of the expressions~(\ref{eq:coeff_functions}) for the coefficient functions in the one-field case. The inverse ratio $\chi^2/h^2$ cannot appear as it would lead to an inconsistency near the bottom of the Higgs potential. 

As before, we solve for torsion, plug the result back into the action and finally move to the Einstein frame by means of the transformation (\ref{eq:Weyl_resc}), where the conformal factor is now given by \Eq \eqref{eq:conformalFactorDilaton}.
After a straightforward computation, we find 
\be
\begin{aligned}
\label{eq:action_gen_Einst_scale}
S_{\rm SI}&=\int \diff^4 x \sqrt{g} \Bigg[\f {M_P^2}{2}\mathring{R} -\f {1} {2\Omega^2} \widetilde \g_{ab} g^{\m\n}\p_\m \varphi_a \p_\n \varphi_b  - \frac{U(h,\c)}{\Omega^4}\Bigg] \ ,
\end{aligned}
\ee
where $\widetilde\g_{ab}~a,b=1,2$, is the metric of the two-dimensional kinetic manifold spanned by $\varphi^a=(\c,h)$; it reads
\be
\widetilde \g_{ab} = \mathcal I_{ab}+\f{4}{\x_\c\c^2\l(G_{vv}G_{aa}-G_{va}^2\r)}\g_{ab}+\f {6}{\x_\c \c^2+\x_h h^2}
\begin{pmatrix}
\x_\c^2\c^2&\x_\c\x_h\c h \\
\x_\c\x_h\c h&\x_h^2h^2
\end{pmatrix}
  \ ,
\ee
with $\mathcal I_{ab}$ the $2\times 2$ identity matrix, and
\be
\begin{aligned}
&\g_{11}=\l(G_{aa}\z^{v\,2}_\c+G_{vv}\z^{a\,2}_\c-2G_{va}\z^v_\c \z^a_\c\r)\c^2 \ ,\\
&\g_{12}=\l(G_{aa}\z_h^v\z^v_\c+G_{vv}\z_h^a\z^a_\c-G_{va}\l(\z_h^v \z^a_\c+\z^v_\c\z^a_h\r)\r)h\c \ ,  \\
&\g_{22}=\l(G_{aa}\z^{v\,2}_h+G_{vv}\z^{a\,2}_h-2G_{va}\z^v_h \z^a_h\r)h^2  \ .
\end{aligned}
\ee
When written in terms of $\c$ and $h$, the kinetic sector of the theory appears to be rather involved. It is also difficult to explicitly identify the physical dilaton when the theory is written this way. Nevertheless, owing to the fact that the kinetic manifold is two-dimensional, we can always diagonalize it (but not necessarily make it canonical). Note that the above construction bears resemblance to theories built on the basis of scale invariance and transverse diffeomorphisms introduced in~\cite{Blas:2011ac}, generalizing the ideas of~\cite{Shaposhnikov:2008xb} in an attempt to have the dilaton be of gravitational origin. These have been further studied, e.g., in~\cite{Karananas:2016grc,Karananas:2016kyt,Casas:2018fum}. We will therefore proceed in the analysis by using the results of the aforementioned works. 

To rid of the kinetic mixing, we first introduce
\be
\label{eq:can_fields_1}
\Phi= \sqrt{\x_\c \c^2+\x_h h^2} \ ,~~~Z= \f{h^2}{\x_\c \c^2} \ ,
\ee
in terms of which the action~(\ref{eq:action_gen_Einst_scale}) becomes
\be
\begin{aligned}
\label{eq:action_gen_Einst_scale_new_variables}
S_{\rm SI}=\int \diff^4 x \sqrt{g} \Bigg[\f {M_P^2}{2}\mathring{R} -\f {M_P^2}{2}\Big ( & \mathcal G_{ZZ}(Z) g^{\mu\nu}\p_\mu Z\p_\nu Z+2\mathcal G_{Z\Phi}(Z) g^{\mu\nu}\p_\mu Z\p_\nu \log(\Phi/M_P) \\
&+\mathcal G_{\Phi\Phi}(Z) g^{\mu\nu} \p_\mu \log(\Phi/M_P) \p_\nu \log(\Phi/M_P) \Big) - \tilde U(Z)\Bigg] \ . 
\end{aligned}
\ee
The various functions appearing in the above depend only on $Z$ and are given by 
\be
\begin{aligned}
&\mathcal G_{ZZ}(Z) =\f{1}{4\x_\c Z(1+\x_h Z)^3}\l( \xi_h^2 Z  \tilde\g_{11} - 2\sqrt{\xi_\chi}\xi_h\sqrt{Z}  \tilde\g_{12}+\xi_\chi  \tilde\g_{22}   \r) \ ,\\
& \mathcal G_{Z\Phi}(Z) = \f{1}{\x_\c (1+\x_h Z)^2}\l(  \xi_h  \tilde\g_{11} + \l( \sqrt{\xi_\chi/Z}-\sqrt{\xi_\chi}\xi_h\sqrt{Z} \r)\tilde\g_{12} +\xi_\chi\tilde\g_{22} \r) \ ,\\
& \mathcal G_{\Phi\Phi}(Z) = \f{1}{\x_\c(1+\x_h Z)}\l(\tilde\g_{11}+2\sqrt{\x_\c Z} \tilde \g_{12}+\x_\c Z \tilde \g_{22} \r) \ , \\
&\tilde U(Z)=\f{\lambda M_P^4}{4\x_\c^2} \l(\f{\f{\alpha}{\lambda} -\x_\c Z}{1+\x_h Z}\r)^2 +\f{\beta M_P^4}{\x_\c^2} \l(\f{1}{1+\x_h Z}\r)^2 \ .
\end{aligned}
\ee
Next, we redefine 
\be
\label{eq:can_fields_2}
\Phi = M_P  {\rm e}^{\f{\tilde \Phi}{M_P}-f(Z)}  \ ,~~~\text{with}~~~f'(Z) = \f {\mathcal G_{Z\Phi}(Z)}{\mathcal G_{\Phi\Phi}(Z)} \ ,
\ee
to obtain 
\be
\begin{aligned}
\label{eq:action_gen_Einst_scale_new_variables_diag}
S_{\rm SI}&=\int \diff^4 x \sqrt{g} \Bigg[\f {M_P^2}{2}\mathring{R} -\f {1}{2}\l( M_P^2\mathcal K(Z) (\p Z)^2 +\mathcal G_{\Phi\Phi}(Z)(\p \tilde \Phi )^2 \r) - \tilde U(Z)\Bigg] \ ,
\end{aligned}
\ee
with 
\be
\mathcal K(Z) = \f{\mathcal G_{ZZ}(Z)\mathcal G_{\Phi\Phi}(Z)-\mathcal G_{Z\Phi}^2(Z)}{G_{\Phi\Phi}(Z)} \ .
\ee

Let us pause for a moment and point out that in terms of the non-canonically normalized $Z$-field, the form of the action can be convenient when studying inflationary physics. The properties of $\mathcal K(Z)$ and, in particular, its pole structure at field values relevant for inflation determine the observable quantities. In turn, this was shown to be related to the geometry of the field space and, more specifically, to its curvature~\cite{Karananas:2016kyt}. After all, we have a two-dimensional manifold, meaning that its scalar curvature fully characterizes the geometry. 

It is clear from the action that $\tilde \Phi$
appears only via its kinetic term so it is manifestly shift-symmetric. This is exactly the NG mode of the nonlinearly realized dilatations. In terms of the original fields the physical dilaton reads
\be
\tilde \Phi = M_P \l [\f 1 2 \log \l(\f {\x_\c \c^2+\x_h h^2} {M_P^2}\r)  +  f\l(\f h \c \r) \r] \ ,
\ee
so that the scale transformations~(\ref{eq:trans_fields}) accompanied by $\chi\mapsto q\chi$ are realized as shifts,
\be
\tilde\Phi~\mapsto~\tilde\Phi + M_P \log q \ .
\ee

Before closing this section, we note that there is no difficulty in making the kinetic term for $Z$ canonical; this is achieved in terms of 
\be
\label{eq:can_fields_3}
\tilde Z = M_P\int \diff Z \sqrt{\mathcal K(Z)} \ , 
\ee
meaning that 
\be
\begin{aligned}
\label{eq:action_gen_Einst_scale_new_variables_canonical}
S_{\rm SI}&=\int \diff^4 x \sqrt{g} \Bigg[\f {M_P^2}{2}\mathring{R} -\f {1}{2}\l( s(\p \tilde Z)^2 +\mathcal G_{\Phi\Phi}(\tilde Z)(\p \tilde \Phi )^2 \r) - \tilde U(\tilde Z) \Bigg] \ ,
\end{aligned}
\ee
with 
\be
s = \text{sign}\l(\mathcal K(Z)\r) \ .
\ee

The theory we constructed here generalizes the metrical Higgs-dilaton model~\cite{Shaposhnikov:2008xb} (see also~\cite{Karananas:2016kyt}) to which it boils down in the limit of vanishing torsion. As mentioned earlier, the Higgs-dilaton model contains many salient features. It connects the early inflationary stage of the Universe to its late-time accelerated expansion \cite{GarciaBellido:2011de}. Besides, the dilaton field $\chi$ may actually be of gravitational origin, if the general covariance of the theory is restricted to transverse coordinate transformations \cite{Karananas:2016grc}. This line of thought is appealing since it again brings us to a situation where there are no new particles of non-gravitational origin, and we saw that at least one new propagating degree of freedom is necessary to realize scale invariance in a phenomenologically viable way. It would be interesting to pursue further these ideas in the EC framework.

\subsection{Inclusion of fermions}
\label{ssec:fermions_scale}

The inclusion of fermions in the scale invariant setting can be done in a straightforward manner, where we employ the same procedure as in~\cite{matter_matters}. The action $S^f_{\rm SI}$ of a massless four-component spinor $\Psi$ comprises its kinetic term for and all possible interactions with torsion. It reads~\cite{Diakonov:2011fs}
\be
\begin{aligned}
\label{eq:ferm_act_1}
&\displaystyle S^{f}_{\rm SI} = \int \diff^4 x\sqrt{g} \Bigg[ \frac{i}{2} \l(\overline{\Psi}\g^\m \mathring{D}_\m\Psi - \overline{\mathring{D}_\m \Psi}\g^\m \Psi \r)
\\ & \qquad\qquad\qquad\qquad+
\l(\z^v_V V_\m +\z^v_A A_\m\r)v^\m + \l(\z^a_V V_\m +\z^a_A A_\m\r)a^\m \Bigg] ,
\end{aligned}
\ee
where $\g^\m=e^\m_A\g^A$, and 
\be 
\mathring{D}_\m = \p_\m +\frac 1 8 \mathring{\omega}_\m^{AB}[\g_A,\g_B] 
\ee
is the torsion-free fermionic covariant derivative. 
In the above $\zeta^{v}_V$, $\zeta^{a}_V$, $\zeta^{v}_A$ and $\zeta^{a}_A$ are arbitrary coefficients, while 
\be 
V_\m = \bar{\Psi} \gamma_\mu \Psi \;, ~~~ A_\m = \bar{\Psi} \gamma_5 \gamma_\mu \Psi
\ee
are the vector and axial fermionic currents, respectively. 

In terms of the equivalent metric theory, the action now includes a set of specific higher-dimensional operators that capture various interactions between and among both scalars and fermions. More specifically, 
\be
\begin{aligned}
\label{eq:action_gen_int-out_fermion}
S &=S_{\rm SI}+\int \diff^4 x \sqrt{g} \Bigg[\f i 2\l(  \overline{\Psi}\g^\m \mathring{D}_\m\Psi -   \overline{\mathring{D}_\m \Psi}\g^\m \Psi\r)\\
&\qquad\qquad\qquad\quad+\frac{1}{\x_\c\c^2}\Big(\mathscr L _ {\c V}^{\rm SI}+\mathscr L _ {h V}^{\rm SI} +\mathscr L _ {\c A}^{\rm SI}+\mathscr L _ {hA}^{\rm SI}+\mathscr L _ {VV}^{\rm SI}+\mathscr L _ {AA}^{\rm SI}+\mathscr L _ {VA}^{\rm SI}\Big)\Bigg] \ ,
\end{aligned}
\ee
where $S_{\rm SI}$ is given above in Eq.~(\ref{eq:action_gen_Einst_scale}), while
\bea
\label{eq:LhS}
& & \mathscr L _{\c V}^{\rm SI} = \frac{G_{aa}\z_\c^v\z^v_V+G_{vv}\z_\c^a\z^a_V-G_{va}(\z^v_V\z_\c^a+\z_\c^v\z^a_V)}{G_{va}^2-G_{vv}G_{aa}}\p_\m \c^2 V^\m \ , \\
& & \mathscr L _{h V}^{\rm SI} = \frac{G_{aa}\z_h^v\z^v_V+G_{vv}\z_h^a\z^a_V-G_{va}(\z^v_V\z_h^a+\z_h^v\z^a_V)}{G_{va}^2-G_{vv}G_{aa}}\p_\m h^2 V^\m \ , \\
& & \mathscr L _{\c A}^{\rm SI} = \frac{G_{aa}\z_\c^v\z^v_A+G_{vv}\z^{a}_\c\z^a_A-G_{va}(\z^a_A\z^{v}_\c+\z^{a}_\c\z^v_A)}{G_{va}^2-G_{vv}G_{aa}}\p_\m\c^2 A^\m \ ,\\
& & \mathscr L _{h A}^{\rm SI} = \frac{G_{aa}\z_h^v\z^v_A+G_{vv}\z^{a}_h\z^a_A-G_{va}(\z^a_A\z^{v}_h+\z^{a}_h\z^v_A)}{G_{va}^2-G_{vv}G_{aa}}\p_\m h^2 A^\m \ ,\\
& & \mathscr L _{VV}^{\rm SI} =\Omega^2\, \frac{G_{aa}(\z_V^v)^2+G_{vv}(\z_V^a)^2-2G_{va}\z_V^v \z_V^a}{2\l(G_{va}^2-G_{vv}G_{aa}\r)}V_\m V^\m \ , \label{eq:LVV_Weyl}\\
& & \mathscr L _{AA}^{\rm SI} =\Omega^2\,\frac{G_{aa}(\z_A^v)^2+G_{vv}(\z_A^a)^2-2G_{va}\z_A^v \z_A^a}{2\l(G_{va}^2-G_{vv}G_{aa}\r)}A_\m A^\m \ , \label{eq:LAA_Weyl}\\
\label{eq:LVA_Weyl}
& & \mathscr L _{VA}^{\rm SI} = \Omega^2\,\frac{G_{aa}\z^v_V\z^v_A+G_{vv}\z^a_V\z^a_A-G_{va}(\z^a_V\z^v_A+\z^v_V\z^a_A)}{G_{va}^2-G_{vv} G_{aa}} V_\m A^\m \ .
\eea
We observe that our result is fully analogous to the findings of \cite{matter_matters}. The only difference is that the presence of two scalar fields leads to two copies of the coupling between a scalar and a fermionic current. Moreover, we note that there is no difficulty in writing the above in terms of the kinetically-decoupled fields $\tilde Z$ and $\tilde \Phi$---this is achieved by using Eqs.~(\ref{eq:can_fields_1}),~(\ref{eq:can_fields_2}) and~(\ref{eq:can_fields_3}). Since the resulting expressions do not offer any new information, we leave this computation to the invested reader. In summary, Eqs.~(\ref{eq:LhS})-(\ref{eq:LVA_Weyl}) show that in principle the fermionic terms can influence the dynamics of the scalar fields as described in Sec.~\ref{ssec:HD}, provided the couplings $\zeta^{v}_V$, $\zeta^{a}_V$, $\zeta^{v}_A$ and $\zeta^{a}_A$ are large enough (see \eg \cite{Shaposhnikov:2020gts}). If the couplings are sufficiently small, however, the previous analysis remains valid even in the presence of fermions.

\section{Weyl invariance}
\label{sec:weyl_inv}

\subsection{General action}

We now attempt to couple the Higgs field to EC gravity in a Weyl-symmetric manner. In other words, we require the theory should be invariant under gauged dilatations. 
Let us briefly outline the procedure of gauging the theory as we would for a non-spacetime symmetry. Our starting point obviously is the globally scale-invariant action~(\ref{eq:action_general_unitary_scale}). To account for the inhomogeneous pieces that now appear in the derivatives of the fields with nontrivial scaling dimensions $d_i$,
we would promote the partial derivatives to covariant ones
\be
\p_\m~\mapsto~D_\m = \p_\m -d_i W_\m \ ,
\ee
by introducing the Weyl vector $W_\m$, transforming as
\be
\label{eq:Weyl_vector}
W_\m~\mapsto~W_\m + q^{-1}\p_\m q \ ,
\ee
where now $q=q(x)$. We would also supplement the action with the appropriate kinetic term for the Weyl gauge field, $(\p_\m W_\n-\p_\n W_\m)^2$, which is manifestly gauge-invariant and moreover satisfies the requirement of being of mass dimension 4. The resulting theory would exhibit Weyl invariance by construction, but at the expense of having new propagating degrees of freedom. 

Although what we just presented is of course an acceptable thing to do, it is neither unique and certainly not the most economic. When it comes to localizing spacetime symmetries, the introduction of compensating gauge fields may turn out to be unnecessary, in the sense that their role can be played by curvature and/or torsion. 
Let us explain why this can be the case to start with. The transformation properties of the fields are still given by Eq.~(\ref{eq:trans_fields}), but since now $q=q(x)$, certain geometric objects---composed out of $e$ and $\omega$---do not transform covariantly but rather pick up inhomogeneous pieces under a Weyl rescaling. For our purposes here we note that the Ricci scalar and the torsion vector shift by derivatives of the scale function as
\bea
\label{eq:Weyl_trans_R}
& & \mathring R~\mapsto~q^2\mathring R + 6 q \square q -12 \l(\p_\m q\r)^2 \ ,\\
\label{eq:Weyl_trans_v}
& & v_\m~\mapsto~v_\m +3q^{-1}\p_\m q \ ,
\eea
with $\square=g^{\m\n}\nabla_\m \nabla_\n$ the covariant d’Alembertian. 

In other words, a theory may be made invariant under Weyl transformations without introducing new degrees of freedom, simply by coupling it appropriately to gravity.\footnote{For more details we refer the interested reader to~\cite{Iorio:1996ad}. This procedure has been systematized both in the absence and the presence of torsion in~\cite{Karananas:2015eha}.} In the EC formulation we can construct covariant derivatives for the fields by letting the torsion vector play the role of an effective Weyl gauge field, as is clear from Eqs.~(\ref{eq:Weyl_vector}) and~(\ref{eq:Weyl_trans_v}). We shall employ this approach, and consequently set
\be
W_\m = \f 1 3 v_\m \ .
\ee
Since torsion is not propagating, $v_\m$ will eventually be related to the derivatives of the fields. In addition, due to the Weyl  redundancy, one scalar degree of freedom is  spurious, meaning that the spectrum of the theory comprises the massless graviton plus a single scalar field. As stated in the introduction, in this context Weyl symmetry is merely a means toward further constraining the action. 

It is a straightforward exercise to write down the scalar-gravity part of the most general Weyl-invariant action involving the Higgs $h$ and another scalar field $\c$, which satisfies conditions~\emph{i)---iii)} from Sec.~\ref{ssec:criteria}. By virtue of the transformation properties of the various quantities, we find that this reads
\be
\begin{aligned}
\label{eq:action_WI_1}
S_{\rm WI} &= \int \diff^4 x \sqrt{g} \Bigg [ (\x_\c\c^2+\x_h h^2)F +(\z_\c\c^2+\z_h h^2)\tilde F - \frac {( D^W_\m  \c)^2}{2}- \frac {( D^W_\m  h)^2}{2} -U(h,\c) \\
&\qquad\qquad\qquad\qquad~~\,+ \f{\x_\c \c^2}{2}\Big(\mathbf{g}_{aa}	a_\mu a^\mu +\mathbf{g}_{\tau\tau}\tau_{\a\b\g} \tau^{\a\b\g} + \tilde{\mathbf{g}}_{\tau\tau}\epsilon^{\m \n \rho \sigma} \tau_{\lambda\m\n} \tau^\lambda_{~\rho\sigma}\Big)+  \tilde J^a_\m a^\m \Bigg]\ ,
\end{aligned}
\ee
where we introduced the Weyl covariant derivative
\be
D^W_\m = \p_\m + \f {1} {3} v_\m \ ,
\ee
and the potential $U(h,\c)$ was defined in Eq.~(\ref{eq:pot}). The difference as compared to the scale-invariant action \eqref{eq:action_general_unitary_scale} is that now the couplings to $v_\m$ no longer appear on their own, but only as part of the curvature and covariant derivatives. The same remains true if we include a complex scalar field or fermions in the theory. In the Weyl-invariant case, the couplings of their currents to $v_\m$ are no longer independent: only the interactions with the axial vector $a_\m$ remain free, and, consequently, the number of a priori undetermined parameters is cut in half.

As before, dimensional analysis dictates that the coefficient functions depend on the ratio of the fields, i.e. $\mathbf{g}_{ij}=\mathbf{g}_{ij}(h/\c)$,  while the current that couples to the pseudo-vector torsion reads
 \be
\tilde J^{a}_\m =  \tilde\z^{a}_\c  D^W_\m \c^2+\tilde\z^{a}_h  D^W_\m h^2\ .
\ee
We find it convenient to recast the action in the following form
\be
\begin{aligned} \label{eq:action_weyl_decomposed}
S_{\rm WI} &= \int d^4 x \sqrt g \Bigg[  \frac{\x_h h^2+\x_\c\c^2}{2} \mathring{R} -\f {(\p_\m h)^2}{2} -\f {(\p_\m \c)^2}{2}- U(h,\c) \\
&-\f 1 6\l[ (1+6\x_h)\p_\m h^2 +(1+6\x_\c)\p_\m\c^2 \r]v^\m -\f {1}{18} \l[ (1+6\x_h) h^2 +(1+6\x_\c)\c^2\r]v_\m v^\m \\
&+\l[ (\z_h+\tilde\z^a_h)\p_\m h^2+(\z_\c+\tilde\z^a_\c)\p_\m \c^2 \r]a^\m + \f 1 3 \l[ (2\z_h+\tilde\z^a_h)h^2+(2\z_\c+\tilde\z^a_\c)\c^2 \r]a^\m v_\m \\
&+\f{\x_\c \c^2}{2}\l(G_{aa}a_\mu a^\mu+G_{\tau\tau}\tau_{\a\b\g} \tau^{\a\b\g} + \tilde{G}_{\tau\tau} \epsilon^{\m \n \rho \sigma} \tau_{\lambda\m\n} \tau^\lambda_{~\rho\sigma}\r)\Bigg] \ ,
\end{aligned}
\ee
where the shifted functions $G_{aa},G_{\tau\tau},\tilde G_{\tau\tau}$ can be read from Eqs.~(\ref{eq:coeff_functions}) and (\ref{eq:g_G}) upon replacing the Planck mass with $\x_\c\c^2$:
\be
\mathbf{g}_{aa}=G_{aa}-\dfrac{\x_h h^2+\x_\c\c^2}{24 \x_\c\c^2}  \ ,~~\mathbf{g}_{\tau\tau}=G_{\tau\tau}-\dfrac{\x_h h^2+\x_\c\c^2}{2 \x_\c\c^2}  \ ,~~\tilde{\mathbf{g}}_{\tau\tau}=\tilde G_{\tau\tau}+\frac{\z_h h^2 +\z_\c\c^2 }{\x_\c\c^2} \ . 
\ee

The analysis of the theory (\ref{eq:action_weyl_decomposed}) can be greatly simplified once we note that only one scalar degree of freedom is dynamical---as we said before, this is an aftermath of the Weyl redundancy of $S_{\rm WI}$. Hence, without loss of generality one can set $\x_\c \c^2=M_P^2$. Dropping also $\tau$ that in any case vanishes \cite{matter_matters}, we find 
\be
\begin{aligned} \label{eq:action_weyl_final}
S_{\rm WI} &= \int d^4 x \sqrt g \Bigg[  \frac{M_P^2+\x_h h^2}{2} \mathring{R} -\f {(\p_\m h)^2}{2}- U(h) \\
&+\f{M_P^2}{2}\l(G_{vv} v_\m v^\m+2G_{va} a^\m v_\m+G_{aa}a_\mu a^\mu\r)+\z^v_h\p_\m h^2 v^\m +\z^a_h\p_\m h^2a^\m \Bigg] \ ,
\end{aligned}
\ee
where now
\be
\label{eq:coef_map}
G_{vv} = c_{vv}\l(1+\f{\x_{vv}h^2}{M_P^2}\r)\ ,~~~G_{va} =c_{va}\l(1+\f{\x_{va}h^2}{M_P^2}\r) \ , ~~~ G_{aa} =c_{aa}\l(1+\f{\x_{aa}h^2}{M_P^2}\r)
\ee
with 
\be \label{eq:coefficientMappingWeyl}
c_{vv}= -\f{1+6\x_\c}{9\x_\c} \ ,~~~\x_{vv}=\f{\x_\c(1+6\x_h)}{1+6\x_\c} \ ,~~~ c_{va}= \f{(2\z_\c+\tilde\z^{a}_\c)}{3\x_\c}\ ,~~~\x_{va}=\f{\x_\c(2\z_h+\tilde\z^a_h)}{2\z_\c+\tilde\z^{a}_\c}\ ,
\ee
and 
\be \label{eq:coefficientMappingWeyl2}
\z^v_h = -\f{1+6\x_h}{6}\ ,~~~\z^a_h = \z_h+\tilde\z^a_h \ . 
\ee
The function $G_{aa}$ is unchanged as compared to \Eq \eqref{eq:action_weyl_decomposed}. 

Let us count the number of free parameters in the action \eqref{eq:action_weyl_final}. We find 7 independent combinations, namely $\x_h$, $\x_\c$, $\z_h$, $\tilde\z^a_h$, the sum 
$2\z_\c+\tilde\z^{a}_\c$ as well as the two constants $c_{aa}$ and $\x_{aa}$. Thus, we have reduced the number of free parameters by $2$ as compared to the model \eqref{eq:action_real_scalar}.

For completeness we present here the equivalent metric theory in the Einstein frame. It reads
\be
\begin{aligned}
\label{eq:action_gen_int-out_Weyl}
S_{\rm WI} &=\int \diff^4 x \sqrt{g} \Bigg[ \f {M_P^2}{2}\mathring{R} -  \f{K(h)}{2}(\p_\m h)^2 - \frac{U(h)}{\Omega^4}\Bigg] \ ,
\end{aligned}
\ee
where $\Omega$ is given in Eq.~(\ref{eq:coeff_functions}) and $K(h)$ is given by (cf. Eq.~(\ref{eq:kh}))
\be
\label{eq:kh_W}
K(h) =\frac{1}{\Omega^2}\Bigg(1+ \f{4h^2}{M_P^2}\l(\frac{G_{aa}\z_h^{v\,2}+G_{vv}\z_h^{a\,2}-2G_{va}\z_h^v \z_h^a}{G_{vv}G_{aa}-G_{va}^2}\r)+ \frac{6\x_h^2h^2}{M_P^2\Omega^2}\Bigg) \ .
\ee
Instead of eliminating the spurious field by brute force, we could have followed the procedure of the previous section. Once we move to the Einstein frame, it is straightforward to write down the kinetic matrix. As it turns out, it only has one nonzero eigenvalue; clearly, this is a manifestation of the fact that one field is nondynamical. 

For a general choice of the couplings, the kinetic term \eqref{eq:kh_W} may not be positive definite at all field values. Nevertheless, since $K(h) \rightarrow 1$ for $h\rightarrow 0$, there always is a range of sufficiently small $h$ for which the theory is consistent. This situation is fully analogous to our findings in the more general model \eqref{eq:action_real_scalar} (see \cite{matter_matters}). We would also like to remark that it is possible to choose the opposite sign for the kinetic term of $\chi$ in Eq.~\eqref{eq:action_WI_1}. Since the dilaton does not propagate, this also results in a consistent theory. As for the equivalent metric theory, it leads to  the replacement $1+6\x_\c \mapsto -1+6\x_\c$ in the kinetic term \eqref{eq:kh_W} (see \Eq \eqref{eq:coefficientMappingWeyl}).

\subsection{A limiting case: ``Weyl-Palatini'' gravity}

The EC theory with the nonminimally coupled scalar field reviewed in Sec.~\ref{sec:recap} contains the Palatini limit. As discussed in~\cite{matter_matters}, it is achieved by including in the gravitational action only the Ricci scalar and a nonminimal coupling to it. In \Eq \eqref{eq:action_real_scalar_F}, it corresponds to setting
\be
\label{eq:WP_choice_1}
\tilde{\Omega} = z_v = z_a = \mathbf{g}_{vv} = \mathbf{g}_{va} = \mathbf{g}_{aa} = \mathbf{g}_{\tau\tau} = \tilde{\mathbf{g}}_{\tau\tau} = 0 \,.
\ee
For this choice of parameters, the second and third terms in Eq.~(\ref{eq:kh}) cancel each other out. Now one can ask what the analog of the Palatini limit is for the Weyl-invariant theory~(\ref{eq:action_WI_1}), which is a constrained version of the one-field model~(\ref{eq:action_real_scalar_F}). As in the general case, we remove all terms from the gravitational action apart from the Ricci scalar and nonminimal couplings to it. This leads to
\be
\label{eq:WP_choice_2}
\z_h=\z_\c =  \tilde\z^{a}_h =  \tilde\z^{a}_\c = \mathbf{g}_{aa} = \mathbf{g}_{\tau\tau} = \tilde{\mathbf{g}}_{\tau\tau} = 0 \ .
\ee
Plugging the above into Eq.~(\ref{eq:action_weyl_decomposed}), we obtain the following action
\be
\begin{aligned} 
\label{eq:action_weyl_Palatini}
S_{\rm WP} &= \int d^4 x \sqrt g \Bigg[  \frac{\x_h h^2+\x_\c\c^2}{2} \mathring{R} -\f {(\p_\m h)^2}{2} -\f {(\p_\m \c)^2}{2}- U(h,\c) \\
&-\f 1 6\l[ (1+6\x_h)\p_\m h^2 +(1+6\x_\c)\p_\m\c^2 \r]v^\m -\f {1}{18} \l[ (1+6\x_h) h^2 +(1+6\x_\c)\c^2\r]v_\m v^\m \\
&+\l(\x_h h^2+\x_\c\c^2\r) \l(\frac{1}{48} a_\mu a^\mu+\frac{1}{4}\tau_{\a\b\g} \tau^{\a\b\g} - \frac{1}{2} \epsilon^{\m \n \rho \sigma} \tau_{\lambda\m\n} \tau^\lambda_{~\rho\sigma}\r) \Bigg] \;.
\end{aligned}
\ee
As the original Higgs-dilaton model, the scalar-gravity sector of the~\emph{Weyl-Palatini theory}~(\ref{eq:action_weyl_Palatini}) contains 2 independent parameters---the nonminimal couplings $\xi_h$ and $\xi_\chi$ of the Higgs and dilaton fields. In the Einstein frame and in the gauge $\xi_\chi\chi^2=M_P^2$, the action becomes of the form~(\ref{eq:action_gen_int-out_Weyl}), where the kinetic function is given by
\be 
\label{eq:WP_K}
K(h)=\frac{1}{\Omega^2}\left( 1-\frac{\x_\c (1+6\xi_h)^2h^2}{(1+6\xi_\chi)M_P^2+\x_\c(1+6\xi_h)h^2}+\frac{6\xi_h^2h^2}{M_P^2\Omega^2}\right) \;.
\ee

Let us comment on one interesting choice of parameters in the Weyl-Palatini theory, namely, $\xi_h=-1/6$. As is evident from Eq.~\eqref{eq:action_weyl_Palatini}, torsion is not sourced in this case. Hence, the theory becomes identical to its counterpart in the metric formulation, where one assumes a priori vanishing torsion. Correspondingly, the dependence on $\xi_\c$ drops out, and the kinetic term reads
\be 
K(h)=\frac{1}{\left(1-\frac{h^2}{6M_P^2}\right)^2} \ .
\ee
We observe that the kinetic sector of the theory is identical to the one of the $\alpha$-attractors (see, e.g., \cite{Kallosh:2015lwa}). Thus, one can wonder whether inflation can be realized in this model. The answer turns out to be negative: due to the presence of the conformal factor in the potential part in Eq.~(\ref{eq:action_gen_int-out_Weyl}), it cannot support slow roll.

Notwithstanding the above, the Weyl-Palatini theory generically represents an economic and phenomenologically viable generalization of the textbook example of a scalar field conformally coupled to gravity. Indeed, many choices of parameters are known to lead to successful inflation in agreement with measurements~\cite{Langvik:2020nrs, Shaposhnikov:2020gts}. We leave a study of observational consequences of the theory for future work.

\subsection{Inclusion of fermions}

Coupling fermionic matter in the Weyl invariant theory can be done in almost the same way as in the scale invariant theory. The only difference is that now the fermionic action cannot contain a mixing term between the torsion vector $v_\m$ and the axial current $A_\m$. The reason is that the latter may not be conserved, even classically, and so a term $A_\m v^\m$ would break the gauge invariance \eqref{eq:Weyl_trans_v}. Consequently, the Weyl-invariant fermionic action reads 
\be
\begin{aligned}
\label{eq:ferm_act_2}
&\displaystyle S^{f}_{\rm WI} = \int \diff^4 x\sqrt{g} \Bigg[ \frac{i}{2} \l(\overline{\Psi}\g^\m \mathring{D}_\m\Psi - \overline{\mathring{D}_\m \Psi}\g^\m \Psi \r)
+
\z^v_V V_\m v^\m + \l(\z^a_V V_\m +\z^a_A A_\m\r)a^\m \Bigg] ,
\end{aligned}
\ee
i.e. it is the same as Eq.~(\ref{eq:ferm_act_1}) for $\z^v_A=0$.

Once torsion is rid of and we set $\x_\c \c^2=M_P^2$, the action boils down to 
\be
\begin{aligned}
\label{eq:action_gen_int-out-Weyl}
S &=S_{\rm WI}+\int \diff^4 x \sqrt{g} \Bigg[\f i 2\l(  \overline{\Psi}\g^\m \mathring{D}_\m\Psi -   \overline{\mathring{D}_\m \Psi}\g^\m \Psi\r)\\
&\qquad\qquad\qquad\quad+\frac{1}{M_P^2}\Big(\mathscr L _ {h V}^{\rm WI} +\mathscr L _ {hA}^{\rm WI}+\mathscr L _ {VV}^{\rm WI}+\mathscr L _ {AA}^{\rm WI}+\mathscr L _ {VA}^{\rm WI}\Big)\Bigg] \ ,
\end{aligned}
\ee
where $S_{\rm WI}$ was presented in~(\ref{eq:action_gen_int-out_Weyl}),  and 
\bea
\label{eq:LhSW}
& & \mathscr L _{h V}^{\rm WI} = \frac{G_{aa}\z_h^v\z^v_V+G_{vv}\z_h^a\z^a_V-G_{va}(\z^v_V\z_h^a+\z_h^v\z^a_V)}{G_{va}^2-G_{vv}G_{aa}}\p_\m h^2 V^\m \ , \\
& & \mathscr L _{h A}^{\rm WI} = \frac{G_{vv}\z^{a}_h\z^a_A-G_{va}\z^a_A\z^{v}_h}{G_{va}^2-G_{vv}G_{aa}}\p_\m h^2 A^\m \ ,\\
& & \mathscr L _{VV}^{\rm WI} =\Omega^2\, \frac{G_{aa}(\z_V^v)^2+G_{vv}(\z_V^a)^2-2G_{va}\z_V^v \z_V^a}{2\l(G_{va}^2-G_{vv}G_{aa}\r)}V_\m V^\m \ , \label{eq:LVV_scale}\\
& & \mathscr L _{AA}^{\rm WI} =\Omega^2\,\frac{G_{vv}(\z_A^a)^2}{2\l(G_{va}^2-G_{vv}G_{aa}\r)}A_\m A^\m \ , \label{eq:LAA_scale}\\
\label{eq:LVA_scale}
& & \mathscr L _{VA}^{\rm WI} = \Omega^2\,\frac{G_{vv}\z^a_V\z^a_A-G_{va}\z^v_V\z^a_A}{G_{va}^2-G_{vv} G_{aa}} V_\m A^\m \ .
\eea
Eqs.~\eqref{eq:LhSW}-\eqref{eq:LVA_scale} correspond to our findings of \cite{matter_matters} for the special case $\z^v_A=0$. Note, however, that the various coefficients and functions are not free in the Weyl-invariant setting, see Eqs.~(\ref{eq:coef_map})-(\ref{eq:coefficientMappingWeyl2}). Regarding the physical implications of fermions, the same conclusions hold as in Sec.~\ref{ssec:fermions_scale}. If the couplings $\z^v_V$, $\z^a_V$ and $\z^a_A$ are chosen to be small enough, they do not influence the dynamics of the scalar fields.

At this point we can go a step further and also discuss what is the dynamics of fermions in the Weyl-Palatini limit. This corresponds to considering the action~(\ref{eq:action_gen_int-out-Weyl}) with the choice of parameters~(\ref{eq:WP_choice_2}), supplemented by
\be
\z^v_V= \z^a_V = 0\ , ~~~ \z^a_A =  - \f 1 8 \ ,
\ee
such that torsion-fermion interactions emerge only from the (fermionic) covariant derivative (see also \cite{matter_matters}). We find that the only interaction that survives is the one between the axial fermionic current:
\be
\mathscr L _ {h V}^{\rm WP}=\mathscr L _ {h A}^{\rm WP}=\mathscr L _ {VV}^{\rm WP} = \mathscr L _ {VA}^{\rm WP}=0 \ ,~~~\mathscr L _ {AA}^{WP} = -\f {3}{16} A_\m A^\m \ .
\ee
In the Einstein-Cartan theory without Weyl-invariance, one finds the same result in the Palatini limit.

\section{Conclusions}
\label{sec:conclusions}

Scale symmetry has interesting applications in both cosmology and particle physics. In the present paper, we considered a scalar field coupled to gravity in the Einstein-Cartan formulation and constructed actions that are invariant under either global or local scale transformations. The resulting models feature nonpropagating torsion and are, in general, phenomenologically viable.

In both cases of scale invariance, in addition to the massless graviton and Higgs boson, there is also an extra scalar degree of freedom, the dilaton. By acquiring a nonvanishing expectation value, this field generates mass scales in the theory. If scale invariance is global, the dilaton is dynamical. Being the Nambu-Goldstone mode of the spontaneously broken symmetry, it couples only derivatively to the rest of the fields. If the spacetime symmetry group of the theory is taken to be volume-preserving coordinate transformations, the scale symmetry is broken in a specific manner: it generates a runaway exponential potential for the dilaton, making it a prime candidate for dynamical dark energy \cite{GarciaBellido:2011de}. As was mentioned in Sec.~\ref{sec:scal_inv}, this allows to accommodate both the inflationary and dark energy dominated epochs of the Universe within a single model, and relate inflationary predictions to the late-time cosmological observations.

\begin{figure}[!t]
        \centering
        \includegraphics[width=0.65\linewidth]{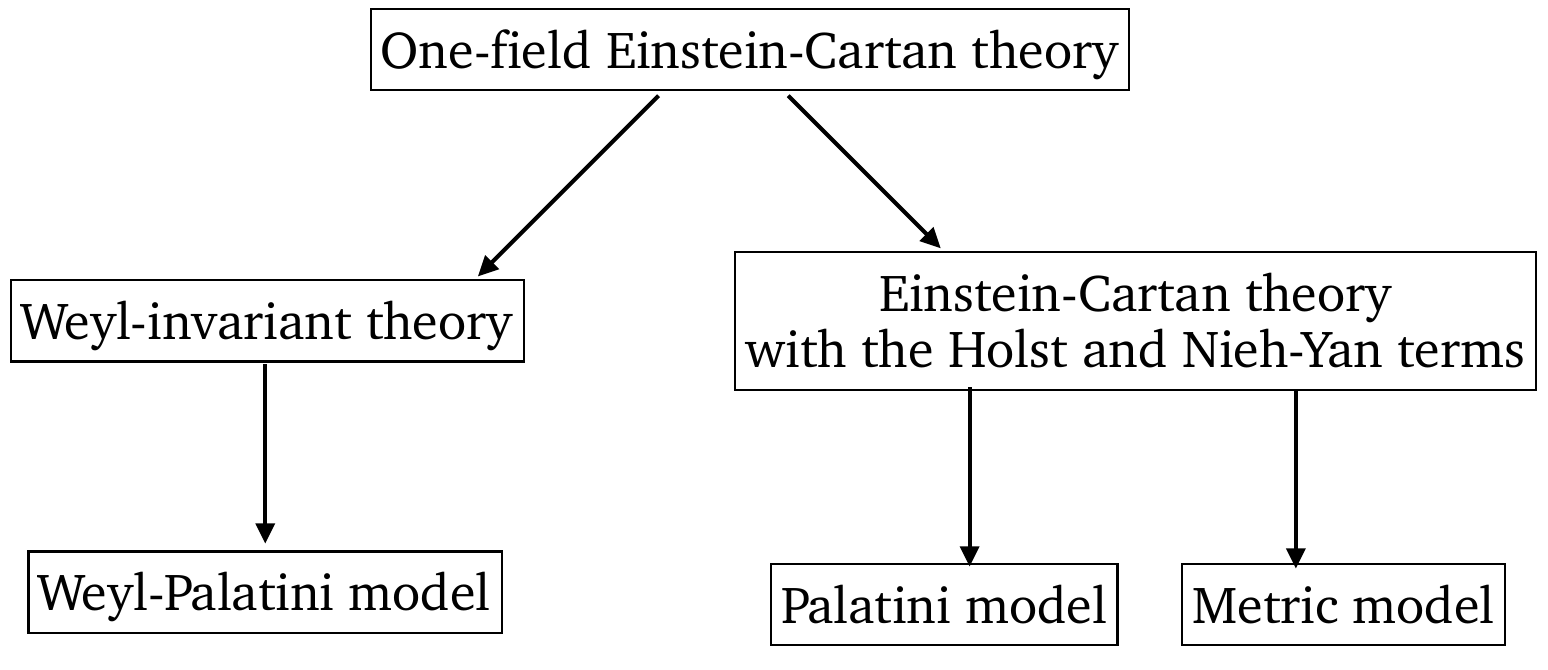}
        \caption{The sub-classes of the general Einstein-Cartan theory with the nonminimally coupled scalar field; see the text for details.}
        \label{fig:table}
\end{figure}

On the other hand, when considering the Weyl invariant generalization, the dilaton is spurious---the gauge redundancy eliminates the field completely. Thus, in the construction that we put forward, the role of Weyl symmetry is to reduce the arbitrariness of the action. By coupling the theory to Einstein-Cartan gravity in a locally scale invariant way, one obtains nontrivial relations between the various coefficient functions. This is a direct generalization of what happens with a conformally coupled scalar in the metric formulation of gravity.

Fig.~\ref{fig:table} summarizes our study of Einstein-Cartan gravity with a nonminimally coupled scalar field. The most general theory contains all terms compatible with the conditions listed in Sec.~\ref{ssec:criteria}. One way to reduce the freedom was explored in~\cite{Shaposhnikov:2020frq} where a subclass containing only the Holst and Nieh-Yan terms was studied. Further limits of the latter theory are the models of Palatini gravity and General Relativity in the commonly-used metric formulation. As discussed, Weyl invariance provides another way of constraining the action. On this route, one can obtain other interesting limiting cases, including the Weyl-invariant analog of the Palatini model. 

Finally, we saw that the inclusion of fermions results in the metric version of the theory  containing a number of higher-dimensional operators describing interactions between the fields and the vector and axial fermionic currents. In large parameter ranges, these additional terms do not influence the dynamics of scalar fields.

\section*{Acknowledgments} 

This work was supported by the ERC-AdG-2015 grant 694896 and by the Swiss National Science Foundation Excellence grant 200020B\_182864. The work of A.~S. was in part supported by the Department of Energy Grant DE-SC0011842.

{\small
\setlength{\bibsep}{1.3pt plus 0.1ex}
\bibliographystyle{utphys}
\bibliography{Nonpropagating_Torsion_SI.bib}
}

\end{document}